\documentstyle[11pt,aasms4,epsf,epsfig]{article}  
\def\la{\lower0.6ex\vbox{\hbox{$ \buildrel{\textstyle 
<}\over{\sim}\ $}}}
\def\ga{\lower0.6ex\vbox{\hbox{$ \buildrel{\textstyle 
>}\over{\sim}\ $}}}

\def\beq{\begin{equation}}
\def\eeq{\end{equation}}

\def\alwaysmath#1{{\ifmmode{#1}\else{$#1$}\fi}}

\def\h1{H\thinspace{$\scriptstyle{\rm I}$}}
\def\d1{D\thinspace{$\scriptstyle{\rm I}$}}
\def\o1{O\thinspace{$\scriptstyle{\rm I}$}}
\def\n1{N\thinspace{$\scriptstyle{\rm I}$}}

\def\etal{{\it et al.}~}

\def\eg{{\it e.g.},~}
\def\3he{$^3$He}
\def\4he{$^4$He}
\def\6li{$^6$Li}
\def\7li{$^7$Li}
\def\3h{$^3$H}

\begin{document}

\vskip 0.7in
 
\begin{center} 

{\Large{\bf ON THE VARIATION OF DEUTERIUM AND OXYGEN ABUNDANCES 
IN THE LOCAL INTERSTELLAR MEDIUM}}
 
\vskip 0.4in
{Gary Steigman}
 
\vskip 0.1in
{\it {Departments of Physics and Astronomy,
The Ohio State University, \\ 
Columbus, OH 43210, USA}}\\
\vskip 1.0in

{\bf Abstract}
\end{center}

The first observations of deuterium and oxygen in the Local Interstellar
Medium (LISM) obtained with the {\it Far Ultraviolet Spectroscopic
Explorer} (FUSE) can be used to search for local abundance variations.  
While the very limited sample of these first data may be consistent with 
no variations, they do offer a hint of {\it anti-correlated} variations 
between D/H and O/H.  If confirmed by more data (which will require 
independently determined, accurate \h1 column densities), these hints 
suggest that observations of interstellar gas within a few kpc of the 
solar neighborhood will reveal clear signs of the evolution of the 
abundance of deuterium from there and then (the Big Bang), to here and 
now (the Local Interstellar Medium of the Galaxy).

\newpage

\noindent

\section{Introduction}

In a series of papers exploring seven lines-of-sight (LOS) in the
Local Interstellar Medium (LISM) the Far Ultraviolet Spectroscopic 
Explorer (FUSE) team (\cite{f02}; \cite{h02}; \cite{k02}; \cite{leh02}; 
\cite{lem02}; \cite{s02}; \cite{w02}) has presented results on the 
column densities of \d1 and \o1 (along with \n1, which will not be 
considered in this paper) but, not of \h1 (due to the absence of 
Ly$\alpha$ within the FUSE spectral range).  These data, which have 
been summarized in Moos \etal (2002), are employed in the analysis 
presented here.  While five of the seven absorbing clouds lie within 
$\sim 80$~pc of the Sun (within the Local Bubble), the other two clouds 
are further away, $\sim 100 - 200$~pc.  Moos \etal (2002) conclude that 
it is likely the deuterium abundance is represented by a single value 
for the five sightlines in the ``near" LISM: D/H $ = 1.52 \times 
10^{-5}$.  While the uncertainty in this mean is $\pm ~0.08 \times 
10^{-5}$, a better measure of the uncertainty might be the weighted
standard deviation which is $\pm ~0.18 \times 10^{-5}$ (Moos, Private
Communication).  It is also claimed that within the Local Bubble the 
\d1/\o1 ratio is constant and they suggest that, as a result, the \o1 
column densities can serve as a proxy for \h1 in the Local Bubble.  
The FUSE team, while cautioning that their results are subject to 
small number statistics, note an increasing dispersion in \d1/\h1 
with increasing distance from the Sun and suggest that this could be 
due to real variations among the LISM deuterium abundances.  However, 
there is no claim of evidence for an anti-correlation between \d1 
and \o1 over the very limited range in metallicity they have explored 
thus far.  

These issues are reconsidered here.  Using the FUSE data (specifically,
Tables 3 \& 4 of Moos \etal 2002), and the same caveats concerning the
limited size of their sample, it is shown that their data is not
inconsistent with small, anti-correlated variations in D/H and O/H.
If so, it becomes problematic to use \o1 as a proxy for the \h1
column densities undetermined from the FUSE data.  The question of
variation or not can only be resolved by more data, especially, to
echo the conclusion of Moos \etal (2002), HST measurements of the \h1
column densities and gas velocity structure.  However, if the variations
suggested here are supported by further data, they offer the promise
of sufficiently large \d1/\o1 variations within a few kpc of the solar 
system that further FUSE data should have no trouble digging the signal 
out of the noise.

In \S2 the FUSE data is used to address the question of variablility
in D/H, O/H, and \d1/\o1 in the LISM.  Having raised the possibility 
of variability, the correlations of D/H and \d1/\o1 with O/H are further 
explored in \S3 where it is suggested that D/H may be anti-correlated 
with O/H.  In \S4 two, likely extreme, forms for such variation are 
considered and compared and the corresponding predicted and FUSE-derived 
abundances of deuterium and oxygen are compared.  In \S5 our conclusions 
and the prospects for future resolution of the issues raised here are 
discussed.

\section{Local Variability?}

Of the seven LOS explored by the FUSE team, two lack estimates of the 
uncertainties in the \h1 column densities and Moos \etal (2002) exclude 
these from their quantitative analyses (except when considering the 
\d1/\o1 column density ratios); the same path is followed here.  In 
Figures 1 -- 4 the various abundances or column density ratios are 
shown versus the \h1 column densities.  Also shown are the data from 
two LOS (towards $\gamma$~Cas and $\delta~$Ori~A) taken from the 
literature (see Table 4 of Moos \etal 2002; \cite{f80}; \cite{m98}; 
\cite{m01}; \cite{j99}).  Those LOS with the largest \h1 column 
densities are also the most distant from the Sun, penetrating the 
Local Bubble and, generally, it is along these LOS that the dispersions 
among the abundance data are greatest.  Given the small sample size 
(five LOS), it may be premature to take the presence of any dispersion 
too seriously.  Nonetheless, it could be a harbinger of real variations 
among the abundances within the LISM.  This latter possibility is 
explored below.

\subsection{Deuterium}

\begin{figure}[ht]
	\centering
	\epsfysize=3.8truein 
\epsfbox{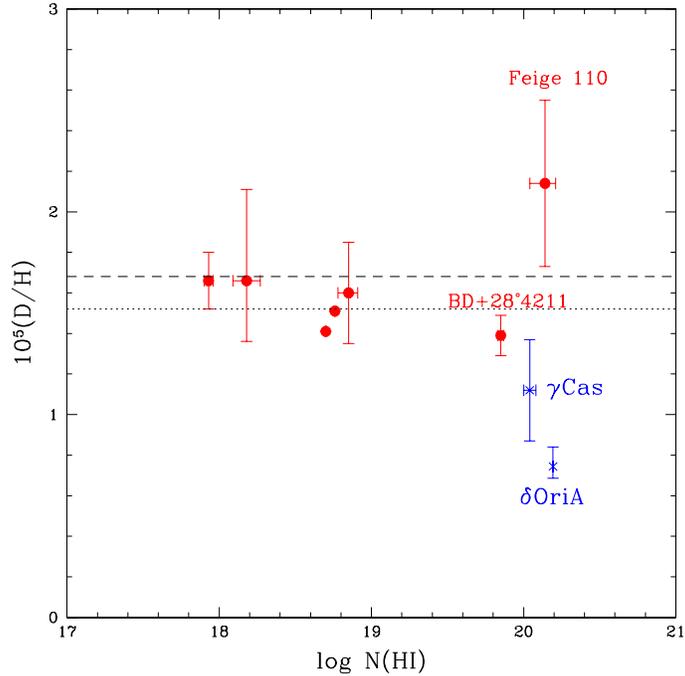}
	\caption{\small{The deuterium abundances along several LOS in 
                the LISM versus the corresponding \h1 column densities.  
                The filled circles are the FUSE data (Moos \etal 2000, 
                Table 3), while the crosses for $\gamma$~Cas and 
                $\delta~$Ori~A are from Copernicus, IUE, IMAPS \& HST 
                (see Table 4 of Moos \etal 2002). The most distant LOS 
                (also the highest \h1 column densities) are identified.
                The dotted line is at the FUSE-determined mean abundance,
                while the dashed line is at the BD~+28$^{\circ}$4211-excluded 
                mean abundance suggested here (see the text).}}
	\label{d/hvsnh}
\end{figure}

The LISM deuterium abundances are plotted in Figure 1.  According to
the data in Moos \etal (2002), the weighted mean deuterium abundance 
is D/H $ = 1.52 \pm 0.08 \times 10^{-5}$.  For this value, the reduced
$\chi^{2}$ ($\chi^{2}$ per degree of freedom) is 1.3, suggesting no
contradiction with the hypothesis that the data are drawn from an 
underlying population with fixed deuterium abundance.  Note, also, 
that three of the five FUSE LOS have deuterium abundances within
1$\sigma$ of this mean value, while the remaining two FUSE LOS have
D/H only slightly more than 1$\sigma$ away (as does $\gamma$~Cas).
In contrast, D/H for $\delta$~Ori~A lies below this mean by more than 
6$\sigma$; unless the uncertaintiess for the column densities along 
this LOS have been seriously underestimated, or affected by unrecognized 
systematic errors, this cloud may have a significantly lower deuterium 
abundance than that in the LISM (\cite{j99}).  Notice that of the five 
FUSE LOS, the one towards BD~+28$^{\circ}$4211, which has the lowest D/H, 
also has the smallest errors, thus tending to dominate the determination 
of the weighted mean abundance.  If, instead, an unweighted mean (for 
all seven of the FUSE LOS) is taken, the mean shifts upward to D/H = 
$1.62 \times 10^{-5}$, moving slightly further away from 
BD~+28$^{\circ}$4211, $\gamma$~Cas, and $\delta$~Ori~A, but slightly 
closer to Feige~110.  On the basis of the FUSE deuterium data alone,
there is no statistical evidence for any variation in the LISM
deuterium abundance.  However, as will be seen next when oxygen 
is considered, there is some evidence that the abundances of
D and/or O towards Feige~110 or BD~+28$^{\circ}$4211 (or both)
may be anomalous.  If the latter LOS is excluded from the estimate
of the weighted mean deuterium abundance, then for the remaining
four FUSE LOS the value increases to D/H = $1.68 \pm 0.11 \times 
10^{-5}$.  This abundance, shown by the dashed line in Figure 
1, provides a very good fit to the data, with a small reduced 
$\chi^{2} = 0.46$ (3 dof).  Notice that three of the five FUSE 
LOS pass within 1$\sigma$ of this value too, but also note that 
BD~+28$^{\circ}$4211 is now nearly 3$\sigma$ away.   

\subsection{Oxygen}

\begin{figure}[ht]
	\centering
	\epsfysize=3.8truein 
\epsfbox{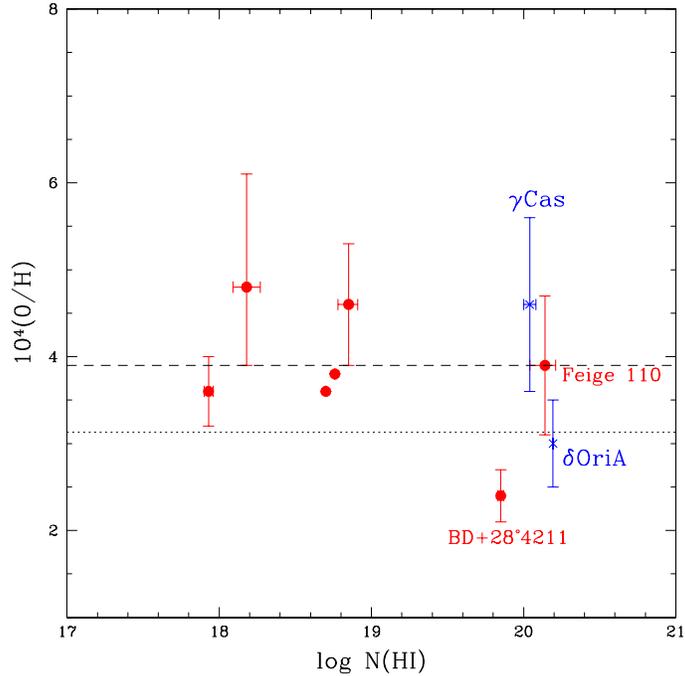}
	\caption{\small{The oxygen abundances along several LOS in the 
                LISM versus the corresponding \h1 column densities.  
                The symbols are as in Figure~\ref{d/hvsnh}.  The most 
                distant LOS are identified.  The dotted line is at the 
                FUSE-determined mean abundance, while the dashed line 
                is at the BD~+28$^{\circ}$4211-excluded mean abundance 
                suggested here (see the text).}}
	\label{o/hvsnh}
\end{figure}

In Figure 2 are plotted the oxygen abundances as a function of the \h1
column densities for the seven FUSE LOS (five with error bars) along
with $\gamma$~Cas and $\delta$~Ori A.  From the data in Table 3 of Moos 
\etal (2002) the weighted mean for the five LOS is O/H = $3.13 \pm 0.21 
\times 10^{-4}$ (this differs slightly from the value, 3.03, quoted in 
Moos \etal 2002, likely due to round-off). What is notable in this case 
is the very large dispersion among the oxygen abundances; the reduced 
$\chi^{2}$, for four degrees of freedom, is 4.0.  In contrast to the 
deuterium abundances, now there is less than a 0.1\% probability that 
these oxygen abundance data have been drawn from an underlying population 
with the weighted mean oxygen abundance.  Notice that of the FUSE 
LOS only Feige~110 is within 1$\sigma$ of this abundance, and that 
BD~+28$^{\circ}$4211 is $\sim 2\sigma$ below the mean.  Although 
the FUSE team identifies Feige~110 as potentially anomalous in D,
they find no evidence for any O-variability for this LOS (see 
\cite{f02}). In fact, if this LOS is removed and the weighted mean 
oxygen abundance is calculated for the four remaining LOS, the mean 
abundance hardly changes at all (from 3.13 to 3.07) while the reduced 
$\chi^{2}$ {\bf increases} to 5.6 (for three degrees of freedom).  It 
would seem that Feige~110 is not the culprit responsible for the dispersion 
among the oxygen abundances.  Indeed, from Figure 2 the smoking gun 
seems to point to BD~+28$^{\circ}$4211 (see, also, \cite{moos02}).  If 
this LOS is excluded instead, the mean oxygen abundance increases to 
O/H = $3.9 \pm 0.3 \times 10^{-4}$ and the reduced $\chi^{2}$ (for 3 
dof) is 0.85.  As may be seen from Figure 2, this is a good fit to the 
limited data set with four of the five FUSE data points along with 
$\gamma$~Cas lying within 1$\sigma$ of this value; the two remaining 
FUSE LOS without error estimates also lie very close to this abundance.  
For this higher oxygen abundance (the dashed line in Figure 2) only 
$\delta$~Ori~A and BD~+28$^{\circ}$4211 are ``outliers''.  Recall 
that the LOS to $\delta$~Ori~A also was a candidate for an anomalously 
low deuterium abundance (see Figure 1), raising the possibility that
the ``problem'' may lie with the \h1 column density determination
(too high?) and not with either the \d1 or \o1 column densities.  
In contrast, for the $\gamma$~Cas LOS, while D/H is somewhat low, 
O/H is at the upper end of the oxygen abundance range, suggesting 
a possible {\it anti-correlation} between D and O along this LOS.  
The same anti-correlation is hinted at for the Feige~110 LOS, but 
for the opposite reason (high-D, somewhat low-O).

\subsection{The \d1 And \o1 Column Densities}

\begin{figure}[ht]
	\centering
	\epsfysize=3.8truein 
\epsfbox{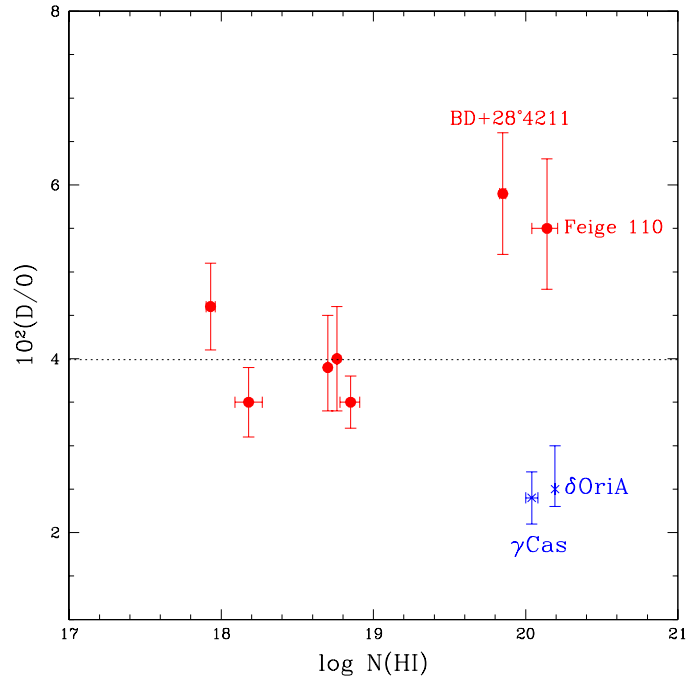}
	\caption{\small{The \d1/\o1 column density ratios along several 
                LOS in the LISM versus the corresponding \h1 column 
                densities.  The symbols are as in Figures~\ref{d/hvsnh} 
                \& \ref{o/hvsnh}.  The dotted line is at the FUSE-determined 
                mean ratio.}}
	\label{d/ovsnh}
\end{figure}

As the Galaxy evolves, incorporating interstellar gas into stars 
and returning stellar processed gas to the ISM, the deuterium 
abundance decreases (deuterium is destroyed in stars), while the 
overall metallicity, in particular the oxygen abundance, increases.  
At some level then, an anticorrelation between D/H and O/H is expected.  
Given the very local sample of interstellar gas in the FUSE data set, 
and the correspondingly very small range in observed abundances, any 
such variations in either D/H or O/H may be hidden by the statistical
errors in the data.  Furthermore, any systematic errors in the non-FUSE 
determinations of the \h1 column densities may either mask -- or 
exaggerate -- any real variations.  To this end, the FUSE-determined 
\d1 and \o1 column densities can play a valuable role.  The consequence 
of charge transfer reactions among H, D, and O in the ISM (Field \& 
Steigman 1971) is to ensure that the \d1/\o1 ratio reflects the {\it 
gas phase} ISM D/O ratio (\eg the D/O ratio modulo any oxygen which 
may be trapped in dust).  This ratio can serve as the canary in the 
coal mine, amplifying any existing, small anticorrelation between D 
and O which might be hidden in the noise of the separate D/H and O/H 
abundance determinations.  To explore this possibility, in Figure 3 
are plotted the \d1/\o1 column density ratios as a function of the 
\h1 column densities.

For all seven of the FUSE LOS the weighted mean \d1/\o1 ratio is
$0.040 \pm 0.002$; this is shown by the dotted line in Figure 3.
However, from Figure 3 it is easy to see, once again, evidence for 
an increasing dispersion (now among the \d1/\o1 ratios) associated 
with the most distant LOS.  Furthermore, only two of the seven 
ratios (two of the five Local Bubble ratios) are within 1$\sigma$ 
of this mean and our two suspect absorbing clouds, those along the 
LOS to BD~+28$^{\circ}$4211 and Feige~110, are between 2$\sigma$ 
and 3$\sigma$ away (the non-FUSE clouds towards $\gamma$~Cas and 
$\delta$~Ori~A are some 3 -- 5~$\sigma$ away).  The reduced $\chi^{2} 
= 3.0$ (for 6 dof) provides no support for the hypothesis that these 
data are drawn from an underlying distribution with a constant \d1/\o1 
ratio.  It may be worth noting that removing Feige~110 from the sample 
only slightly reduces the mean ratio, from 0.040 to 0.039, while the 
reduced $\chi^{2}$ is only slightly reduced from 3.0 (for 6 dof) to 
2.6 (for 5 dof).  If, instead, BD~+28$^{\circ}$4211 is removed, the 
mean is virtually unchanged while there is an improvement in the 
reduced $\chi^{2}$ to 2.0; for 5 dof this still does not provide 
support for the hypothesis of a constant \d1/\o1 ratio.  Either the 
data (FUSE and non-FUSE) is contaminated by larger than estimated 
statistical errors, or by unidentified systematic uncertainties 
(or both) or, the \d1/\o1 ratios are suggesting that there may be 
real abundance variations in D/H and/or O/H between and among the 
nearby and the more distant absorbing clouds.  This latter possibility 
is explored next.

\begin{figure}[ht]
	\centering
	\epsfysize=3.8truein 
\epsfbox{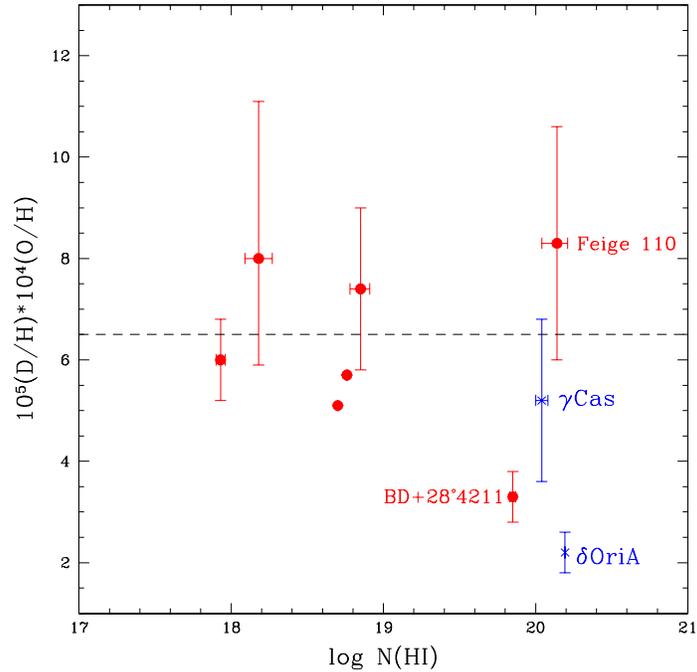}
	\caption{\small{The product of the deuterium and oxygen
                abundances versus the \h1 column densities.  The
                symbols are as in Figures~\ref{d/hvsnh} -- 
                \ref{d/ovsnh}. The dashed line is at the weighted 
                mean for the product (see the text).}}
	\label{dxovsnh}
\end{figure}

If, indeed, at least some of the dispersion in the \d1/\o1 ratios
uncovered above are due to real variations in the deuterium and/or
oxygen abundances, it might be expected that the variations in these 
two abundances should be anticorrelated.  However, while only a small 
amount of gas need be cycled through stars to produce a noticeable 
change in metallicity of the ISM, any observable change in the deuterium 
abundance requires that a significant, and significantly different 
fraction of the gas in some clouds has been processed through stars.  
As a result, it may well be that observable differences exist among 
oxygen abundances along different LOS in the LISM, while the changes 
in deuterium abundances are too small to be detected.  Indeed, this is 
suggested by the FUSE results (see \S2.1 and \S2.2) where a constant 
D/H is consistent with the data while a constant O/H is disfavored.  
If, however, the deuterium and oxygen abundances are both varying, 
and they are anti-correlated, then as an example of an extremely 
strong anti-correlation, their product might be nearly constant.  
In Figure 4 the product of the deuterium and oxygen abundances is 
shown (versus the \h1 column densities).  Notice that, with the 
exception of BD~+28$^{\circ}$4211 whose product of abundances is 
low, all the remaining FUSE LOS are in a rather narrow range of 
each other.  It has already been noted that {\bf both} the deuterium 
and the oxygen abundances for BD~+28$^{\circ}$4211 are low, suggesting 
that the culprit might be the \h1 column density determination along 
this LOS.  If so, this would be exacerbated in the product of abundances.  
If BD~+28$^{\circ}$4211 is excluded, the weighted mean for the product 
of $y_{\rm D} \equiv 10^{5}$(D/H) and $y_{\rm O} \equiv 10^{4}$(O/H) 
is $6.5 \pm 0.7$; this is shown by the dashed line in Figure 4.  The 
remaining four FUSE LOS are all within 1$\sigma$ of this value (and 
the two remaining FUSE LOS are close by) and the reduced $\chi^{2} 
= 0.6$ (for 3 dof).  Thus, although a constant D/H is entirely 
consistent with the FUSE data (see \S2.1), there is some evidence 
in the same data for variations in O/H (see \S2.2) which may be 
anticorrelated with small variations in D/H.  Notice that for 
$\gamma$~Cas, which has a deuterium abundance below the mean (see 
Figure 1) and an oxygen abundance at the high end of the range (see 
Figure 2), the product of the two abundances is completely consistent 
with the mean value.  The same is true for Feige~110 which, in contrast, 
has slightly high-D and slightly low-O.  Finally, note that, like 
BD~+28$^{\circ}$4211, $\delta$~Ori~A is low both in D/H and O/H and, 
as a consequence, lies far from the mean of the product of deuterium 
and oxygen abundances.  

\section{Correlations With Oxygen?}

\begin{figure}[ht]
	\centering
	\epsfysize=3.8truein 
\epsfbox{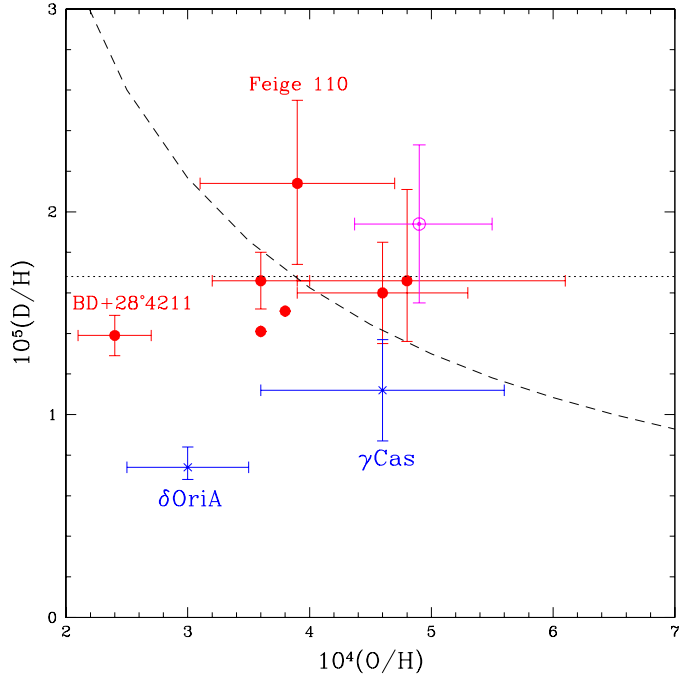}
	\caption{\small{The deuterium abundances along several LOS in 
                the LISM versus the corresponding oxygen abundances.  
                The symbols are as in Figures~\ref{d/hvsnh} -- 
                \ref{dxovsnh}.  The solar symbol is for the solar 
                system (pre-solar nebula) abundances (see the text).  
                The most distant LOS are labelled.  The dotted line 
                is at the revised mean value of the deuterium abundance 
                recommended here (D/H $ = 1.7 \times 10^{-5}$; see \S2.1), 
                while the dashed line shows the D vs. O anti-correlation 
                proposed here ((D/H)(O/H) $ = 6.5 \times 10^{-9}$; see 
                the text and Figure~\ref{dxovsnh}).}}
	\label{dvso}
\end{figure}

If, as suggested above, the FUSE data hints at local abundance
variations which may be anticorrelated between deuterium and
oxygen, these variations should emerge when D/H, \d1/\o1, or
$y_{\rm D}\times y_{\rm O}$ are compared with O/H (unless, 
of course, statistical or systematic errors are responsible 
for the suggested variations).  To this end, in  Figures 5 \& 
6 are shown D/H versus O/H and \d1/\o1 versus O/H respectively.
In these figures, for the purpose of comparison, the solar
system deuterium and oxygen abundances (Geiss \& Gloeckler 
1998, Gloeckler \& Geiss 2000; Allende-Prieto, Lambert, \& 
Asplund 2001) are also included. Notice that considering the
relatively large errors for the solar system (pre-solar nebula)
deuterium abundance (\cite{gg98} \& \cite{gg00}), along with
the lower, revised solar oxygen abundance of Allende-Prieto, 
Lambert, \& Asplund (2001), the solar system abundances are
not at all inconsistent with those found in the 4.6~Gyr younger
gas in the LISM.  Indeed, it should be kept in mind that the
gas phase oxygen abundances may only be {\it lower} limits to
the true ISM oxygen abundance since some oxygen may be tied 
up in dust grains.  If, for example, the suggestion of Esteban 
\etal (2002; see also, Esteban \etal 1998) of an 0.08 dex 
correction for dust were adopted, the mean LISM oxygen abundance 
would increase from the \h1 value of $3.9 \times 10^{-4}$ found 
here, to $4.7 \times 10^{-4}$, in excellent agreement with the 
solar value.  At the same time, it should be noted that the 
photospheric value chosen here (\cite{ala01}; see also~\cite{h01}) 
may only be a {\it lower} bound to the pre-solar nebula abundance 
since over the 4.6~Gyr life of the Sun, some oxygen may have 
settled out of the photosphere.

Figure 5 provides a reflection of the conclusions reached in
\S2 that while the FUSE data may be consistent with a constant
deuterium abundance, they are also not inconsistent with a small
variation  in deuterium abundances which is anticorrelated with 
a similarly small variation in oxygen abundances.  This latter 
option receives further support in Figure 6 where it is clear 
that while a constant \d1/\o1 ratio is incapable of accounting 
for the bulk of the data, the ratios do support a variation in 
oxygen abundance which may either be uncorrelated with any variation 
in D/H (dotted curve) or anticorrelated with a deuterium abundance
variation (dashed curve).

\section{Discussion}

\begin{figure}[ht]
	\centering
	\epsfysize=3.8truein 
\epsfbox{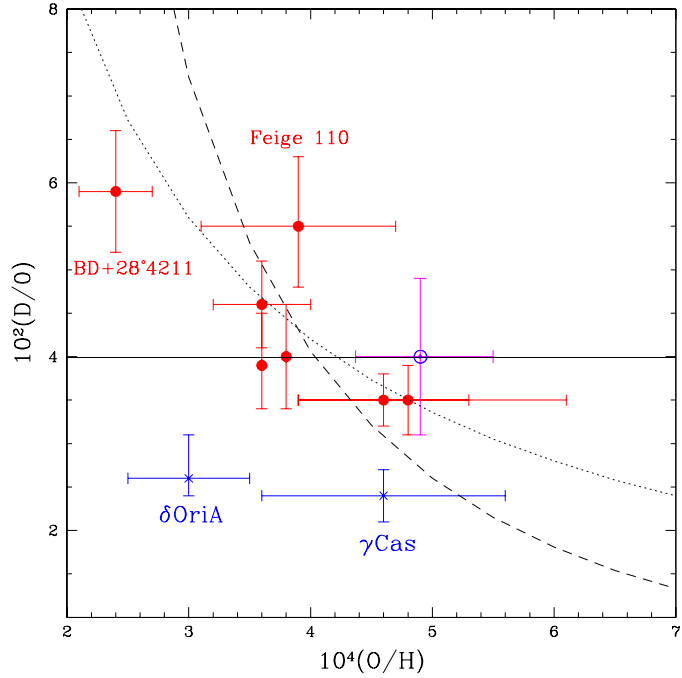}
	\caption{\small{The \d1/\o1 column density ratios along 
                several LOS in the LISM versus the corresponding 
                oxygen abundances.  The symbols are as in Figure~
                \ref{dvso}.  The solid line is at the mean value 
                of \d1/\o1 found in \S2.3, while the dotted line 
                assumes D/H = $1.7 \times 10^{-5}$ is constant, 
                and the dashed line shows the D vs. O anti-correlation 
                suggested here (see the text and Figure~\ref{dxovsnh}).}}
	\label{d/ovso}
\end{figure}

\begin{figure}[ht]
	\centering
	\epsfysize=3.8truein 
\epsfbox{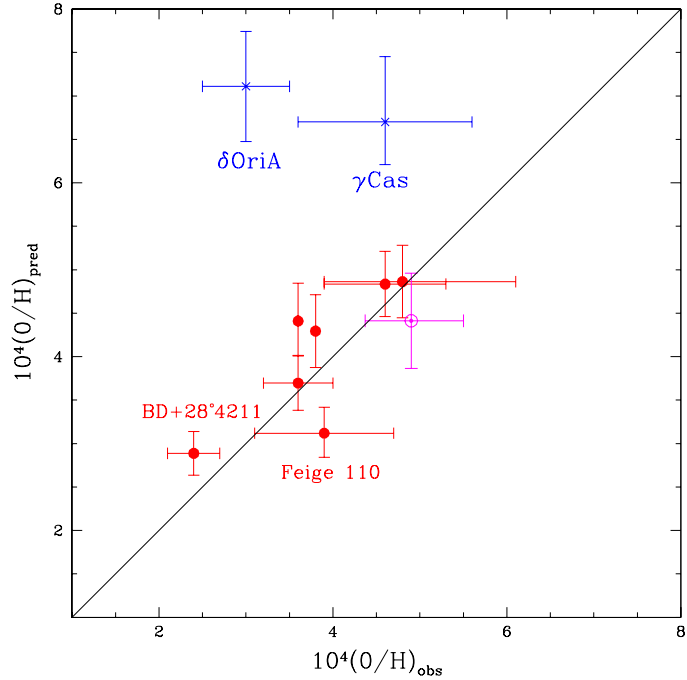}
	\caption{\small{The predicted (vertical) versus observed
                (horizontal) oxygen abundances on the assumption
                of a constant deuterium abundance (eq.~2).  The 
                symbols are as in the other figures.}}
	\label{ovso1}
\end{figure}

\begin{figure}[ht]
	\centering
	\epsfysize=3.8truein 
\epsfbox{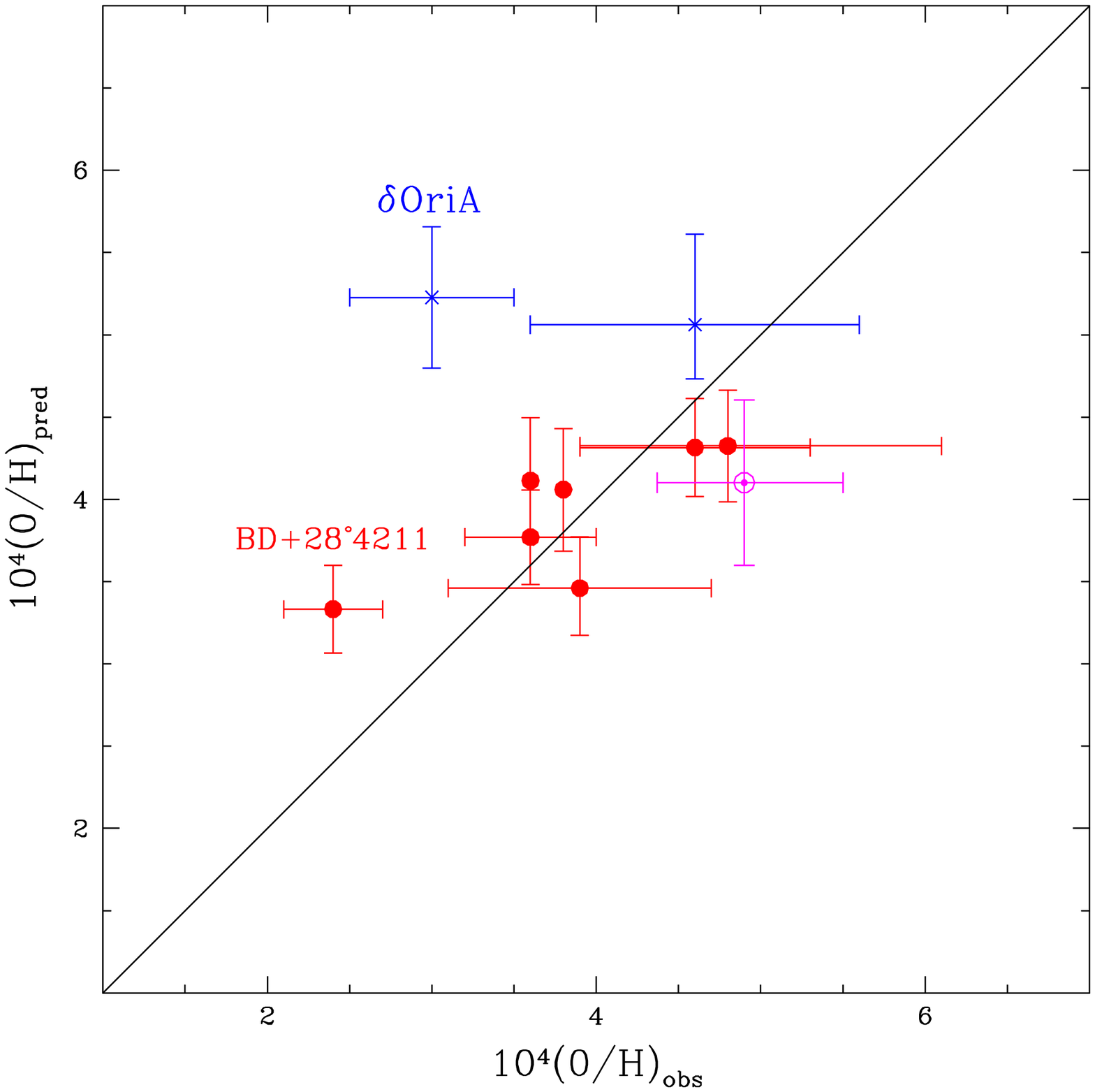}
	\caption{\small{The predicted (vertical) versus observed
                (horizontal) oxygen abundances on the assumption
                that D and O are anticorrelated (eq.~4).  The 
                symbols are as in the other figures.}}
	\label{ovso2}
\end{figure}

Since the FUSE spectral range does not include \h1 (or \d1) Ly$\alpha$,
and the higher lines of the Lyman series lie on the flat part of the
curve of growth for the LOS in the LISM, the FUSE team has relied on 
independent determinations of the \h1 column densities.  Because of 
this limitation, they suggest that it might, instead, be possible to 
use the \o1 column densities as surrogates for the \h1 column densities.  
For example, since
\beq
z \equiv 10^{2}({\rm D}/{\rm O}) = 10y_{\rm D}/y_{\rm O},
\eeq
then {\bf provided that the deuterium abundance is constant}, 
$y_{\rm D} = <y_{\rm D}> = 1.7 \pm 0.1$, 
\beq
y_{\rm O} = 10<y_{\rm D}>/z = {17 \pm 1 \over z}, 
\eeq
so that a measurement of \d1/\o1 ($\propto z$) leads directly to a 
predicted oxygen abundance.  This relation is shown by the dotted 
curve in Figure 6.  This is not at all inconsistent with the FUSE 
data.  In this case a measurement of $z$ leads to a {\it predicted}
oxygen abundance (eq.~2) which may be compared to those derived
from the FUSE (and other) observations.  In Figure 7 is shown the
relation between the currently available observed and predicted
oxygen abundances.

But, a constant deuterium abundance is not required by the data.  
Indeed, it has been seen that the data are also consistent with 
small variations in, along with a rather strong anticorrelation 
between, deuterium and oxygen ($y_{\rm D} \propto 1/y_{\rm O}$).  
In this case,
\beq
z = {10<y_{\rm D}\times y_{\rm O}> \over y_{\rm O}^{2}} = {65 \pm 7 \over 
y_{\rm O}^{2}},
\eeq
so that
\beq
y_{\rm O} = ({{65 \pm 7} \over z})^{1/2}.
\eeq
In this latter case, deuterium will vary along with oxygen so that
\beq
y_{\rm D} = (0.1<y_{\rm D}\times y_{\rm O}>z)^{1/2}.
\eeq

On the assumption that both D and O are varying locally, equations
2 \& 3 may be used, along with the \d1/\o1 column density ratios 
$z$, to {\bf predict} the oxygen and deuterium abundances ($y_{\rm O}$ 
and $y_{\rm D}$).  In Figures 8 \& 9 these predictions are compared 
with the current FUSE (and other) data. Now, neither Feige~110 nor 
$\gamma$~Cas is anomalous and even the solar system values are close 
to those predicted.  The only outliers from these $y_{\rm O}$ vs. 
$z$ and $y_{\rm D}$ vs. $z$ relations are BD~+28$^{\circ}$4211 and 
$\delta$~Ori~A.  A possible source of their apparently anomalous 
abundances is discussed below.

\begin{figure}[ht]
	\centering
	\epsfysize=3.8truein 
\epsfbox{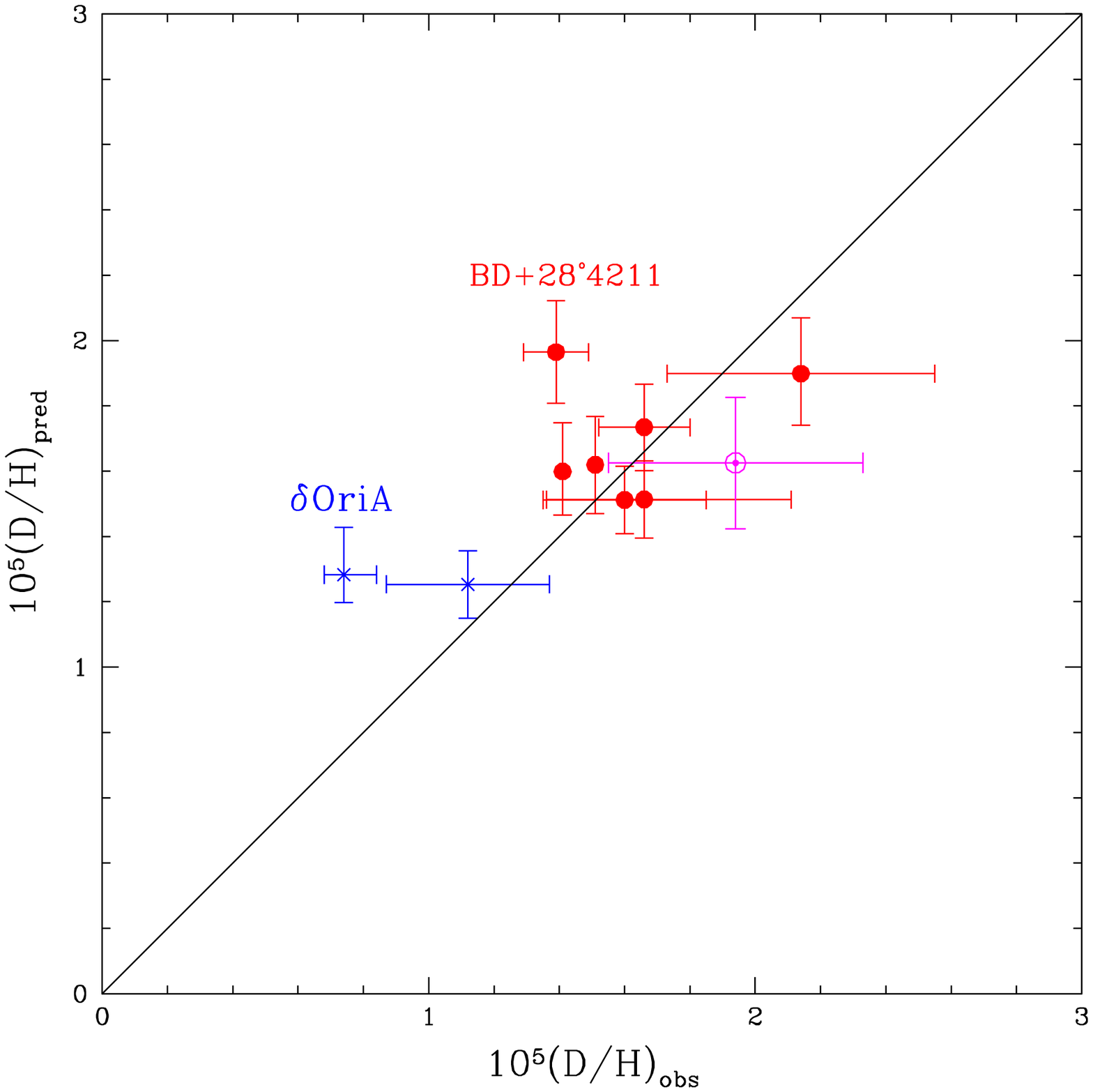}
	\caption{\small{The predicted (vertical) versus observed
                (horizontal) deuterium abundances on the assumption
                that D and O are anticorrelated (eq.~5).  The 
                symbols are as in the other figures.}}
	\label{dvsd}
\end{figure}

On the basis of the current, very limited FUSE data set it is not 
possible to decide between the two options explored here (D varying 
or constant).  To resolve this conundrum will require more data 
{\bf with well-determined \h1 column densities}.  The good news 
though, is that if indeed there are real variations in the currently 
very limited, very local FUSE data sample (as, perhaps, bounded by 
the dotted and dashed curves in Figure 6), future data from within a 
few kpc of the Sun should reveal statistically significant differences 
in the \d1/\o1 column density ratios.

\subsection{BD~+28$^{\circ}$4211 and $\delta$~Ori~A}

The excess dispersion in the FUSE-determined LISM abundances may be 
due to one or more of several possible sources.  The sample is small 
and the statistical errors may have been underestimated.  For some 
column densities along some LOS there may be unidentified systematic 
errors.  Or, there may be real variations in the oxygen and deuterium 
abundances, even for this very local sample.  The latter possibility 
has been explored here and it has been noted that the current data 
cannot exclude this option.  The hypothesis of a -- surprisingly 
strong -- anticorrelation between D and O ($y_{\rm D} \propto 1/y_{\rm O}$; 
see Figure 4) is not at all inconsistent with the FUSE data.  
The only outliers to this anticorrelation are BD~+28$^{\circ}$4211 
and $\delta$~Ori~A (see Figures~4 -- 6).  Along {\bf both} these LOS 
{\bf both} the deuterium and oxygen abundances are low (see also 
Figures~7 -- 9).  Moos \etal (2002) note that BD~+28$^{\circ}$4211 
(as well as Feige~110) has a complex photospheric spectrum (\cite {s02}) 
and that the placement of the continuum, crucial for accurate column 
density determinations, ``was hindered by the complexity of the metal 
lines and the poorly known atomic data for some of the species arising 
in the photospheres of these stars".  It is the case that for both 
BD~+28$^{\circ}$4211 and $\delta$~Ori~A the estimated abundance errors 
are dominated by the errors in the \d1 and \o1 column densities.  
As noted by the referee (Private Communication) in $\delta$~Ori~A, 
\o1 is  determined from a single very weak line, while the component 
structure of the neutral gas toward BD~+28$^{\circ}$4211 is unknown 
and could result in a ``low" N(\o1) or N(\d1) if the b-values vary 
significantly between components (FUSE lacks the spectral resolution 
to define this velocity structure).  If, as suggested here 
BD~+28$^{\circ}$4211 and $\delta$~Ori~A are anomalous, it should be
noted that for the former the deuterium and oxygen abundances can
be reconciled with the FUSE-determined means by increasing N(\d1)
by 1.4$\sigma$ and N(\o1) by 2.4$\sigma$.  For $\delta$~Ori~A, while
the O-abundance is less than 1$\sigma$ from the mean, the D-abundance
differs from the mean by more than 6$\sigma$.  Especially in this
latter case it is unlikely that such a difference, if not reflecting
D-destruction, can be blamed on a statistical fluctuation.

Perhaps, however, the {\it problem} is not with either the \d1 or \o1 
column densities, but with the \h1 column densities along these LOS.  
For BD~+28$^{\circ}$4211, a decrease of only $\sim 0.13$~dex would be 
sufficient to bring the abundances along this LOS into agreement (within 
the remaining statistical uncertainties) with our suggested anticorrelation: 
$y_{\rm D} \times y_{\rm O} = 6.5 \pm 0.7$.  While the same may be true 
for $\delta$~Ori~A, a somewhat larger decrease in N(\h1), $\sim 0.2$~dex, 
would be required.  It must be noted, however, that such large shifts
seem unlikely given the quality of the data and the \h1 column densities 
which are so large that the broad damping wings of Ly-$\alpha$ alpha 
should provide a very good constraint on N(HI).  It should be noted
that such shifts would be more than six times the quoted statistical
errors in the \h1 column densities.  Such large changes would need to
result from systematic errors, although the authors (\cite{j99},
\cite{s02}) have been very careful in their analyses.

To shed new light on these issues it could be of value to reobserve 
these two LOS (and others) with a view to reexamining the \h1, \o1, 
and \d1 column density determinations in order to disentangle true
abundance variations from possible systematic uncertainties.

\section{Conclusions}

The FUSE data have been used to revisit the question of possible
abundance variations in the LISM.  The sample is painfully limited
(seven LOS; only five with \h1 column density determinations with
quoted uncertainties) but, within the statistical errors, the ananlysis
presented here provides a hint of some variations in the local oxygen 
abundance (by the excess dispersion around the mean abundance) which 
may be anticorrelated with some variations in the LISM deuterium abundance.  
This is in contrast to the conclusions of Moos \etal (2002).  Among 
the seven FUSE LOS and the two additional LOS considered by Moos 
\etal (2002), two outliers are identified: BD~+28$^{\circ}$4211 and 
$\delta$~Ori~A.  The former, from the FUSE data set, has the smallest 
statistical errors for the D- and O-abundances, and thus dominates the 
FUSE mean abundance determinations (largely due to the very small error 
adopted for the \h1 column density determination).  When 
this LOS is excluded from the sample, the mean D- and O-abundances 
increase slightly: $<y_{\rm D}> = 1.7 \pm 0.1$, $<y_{\rm O}> = 3.9 \pm 
0.3$.  The remaining FUSE data, while not inconsistent with a constant 
D-abundance in the LISM, still have an unexpectedly large dispersion 
around the mean O-abundance, suggesting that there may be real oxygen 
abundance variations along nearby LOS.  If, indeed, there are variations 
in O/H in the LISM, they might be {\it anti}-correlated with variations 
in D/H since as gas is cycled through stars deuterium is destroyed.  
The FUSE data set is, indeed, not inconsistent with a constant product 
of deuterium and oxygen abundances.  If this anticorrelation is confirmed 
by further data, there is both good news and bad news.  The bad news 
is that as FUSE expands its horizon beyond the LISM, it is unlikely 
that the ratio of \d1 to \o1 column densities ($z \equiv 10^{2}$D/O) 
can serve as a surrogate for independent \h1 column density measurements 
in the determination of D- and O-abundances.  The good news is that 
even within a few kpc of the Sun, based on estimates of the oxygen 
and deuterium abundance gradients in the Galaxy (\cite{mv00}, \cite{cm00}) 
$y_{\rm O}$ and $y_{\rm D}$ will vary sufficiently so that the amplification 
of their ratio, $z$, will result in $z$-variations (\eg by roughly a factor 
two over $\sim 2$~kpc) which will be more easily seen above the background 
of the statistical uncertainties.

It should be noted that even if the rather strong anticorrelation, 
consistent with the current FUSE data set, is confirmed locally, 
such a strong anticorrelation is unlikely to extend to much lower 
oxygen abundances.  Indeed, as pristine gas from the early universe 
begins to be processed through stars, the heavy element abundances, 
oxygen in this case, will quickly increase from their zero primordial 
values before very much gas has been cycled through stars, destroying 
deuterium.  As a result, for a long time (as measured by metallicity) 
the deuterium abundance will not deviate noticeably from its relic 
value, while the oxygen abundance will increase by orders of magnitude 
(the deuterium ``plateau'').  For example, if within the Galaxy a factor 
two lower oxygen abundance (than in the LISM) were accompanied by a 
factor two higher deuterium abundance, the result would be a D-abundance 
indistinguishable from the current estimates of the relic primordial 
D-abundance inferred from observations of gas in high redshift, low 
metallicity QSO Absorption Line Systems (\cite{bt98a};~\cite{bt98b};
~\cite{om01};~\cite{pb01};~\cite{ddm01};~\cite{lev02}).  Indeed, the 
mean LISM D-abundance proposed here, $y_{\rm D} = 1.7 \pm 0.1$, is 
already indistinguishable from that suggested by Pettini \& Bowen 
(2001; PB) for a high redshift (z~$ \sim 2$), low metallicity 
([Si/H]~$ \la -2$) QSOALS: $y_{\rm D}({\rm PB}) = 1.65 \pm 0.35$.  
The deuterium abundances derived from observations of the other 
QSOALS range from $y_{\rm D}({\rm QSOALS}) \approx 2.5$ to 4.0.  
Therefore, it might be anticipated that future FUSE data along LOS 
within a few kpc of the Sun might be capable of mapping the evolution 
of deuterium back to the primordial deuterium plateau, providing a 
valuable complement to the very difficult searches for primordial-D 
in the QSOALS.  

\noindent {\bf Acknowledgments}

I gratefully acknowledge helpful and informative correspondence with 
several members of the FUSE team, in particular Ed Jenkins, Warren Moos, 
Ken Sembach and George Sonneborn.  I also thank the referee for several 
helpful suggestions which have been incorporated in the manuscript.  
This research is supported at The Ohio State University by DOE grant 
DE-AC02-76ER-01545.  Some of this work was done while visiting IAGUSP, 
Brasil and I thank them for their hospitality.


\begin{thebibliography}{}

\bibitem[Allende-Prieto, Lambert \& Asplund 2001]{ala01} Allende-Prieto, 
C., Lambert, D. L., \& Asplund, M. 2001, ApJ, 556, L63

\bibitem[Burles \& Tytler 1998a]{bt98a} Burles, S. \& Tytler, D. 1998a,
ApJ, 499, 699

\bibitem[Burles \& Tytler 1998b]{bt98b} Burles, S. \& Tytler, D. 1998b, 
ApJ, 507, 732

\bibitem[D'Odorico, Dessauges-Zavadsky \& Molaro 2001]{ddm01} D'Odorico, 
S., Dessauges-Zavadsky, M., \& Molaro, P. 2001, A\&A, 368, L21 

\bibitem[Chiappini \& Matteucci 2000]{cm00} Chiappini, C. \& Matteucci, 
F. 2000, Proceedings of IAU Symposium 198, The Light Elements and Their 
Evolution (L. da Silva, M. Spite, and J. R. Medeiros eds.; ASP 
Conference Series) p. 540
 
\bibitem[Esteban \etal 1998]{est98} Esteban, C., Peimbert, M., 
Torres-Peimbert, S., \& Escalante, V. 1998, MNRAS, 295, 401 

\bibitem[Esteban \etal 2002]{est02} Esteban, C., Peimbert, M., 
Torres-Peimbert, S., \& Rodr\' \i guez, M. 2002, preprint 
(astro-ph/0208313)

\bibitem[Field \& Steigman 1971]{fs71} Field. G. B. \& Steigman, G. 
1971, ApJ, 166, 59

\bibitem[Ferlet \etal 1980]{f80} Ferlet, R., Vidal-Madjar, A., Laurent, 
C., \& York, D. G. 1980, ApJ, 242, 576

\bibitem[Friedman \etal 2002]{f02} Friedman, S. D., \etal 2002, ApJS, 
140, 37

\bibitem[Geiss \& Gloeckler]{gg98} Geiss, J., \& Gloeckler, G. 1998, 
Space Sci. Rev., 84, 239  

\bibitem[Gloeckler \& Geiss 2000]{gg00} Gloeckler, G., \& Geiss, J. 
2000, Proceedings of IAU Symposium 198, The Light Elements and Their 
Evolution (L. da Silva, M. Spite, and J. R. Medeiros eds.; ASP 
Conference Series) p. 224 

\bibitem[H\'ebrard \etal 2002]{h02} H\'ebrard, G., \etal 2002, ApJS, 
140, 103

\bibitem[Holweger 2001]{h01} Holweger, H. 2001, in Solar and Galactic
Composition, ed. R. F. Wimmer-Schweingruber, AIP Conf. Proc. 598, 23

\bibitem[Jenkins \etal 1999]{j99} Jenkins, E. B., Tripp, T. M., 
Wozniak, P. A., Sofia, U. J., \& Sonneborn, G. 1999, ApJ, 520, 182

\bibitem[Kruk \etal 2002]{k02} Kruk, J. W., \etal 2002, ApJS, 
140, 19

\bibitem[Lehner \etal 2002]{leh02} Lehner, N., \etal 2002, ApJS, 
140, 81

\bibitem[Lemoine \etal 2002]{lem02} Lemoine, M., \etal 2002, ApJS, 
140, 67

\bibitem[Levshakov \etal 2002]{lev02} Levshakov, S. A., 
Dessauges-Zavadsky, M., D'Odorico, S., \& Molaro, P. 2002, ApJ, 
565, 696

\bibitem[Martins \& Viegas 2000]{mv00} Martins, L. P. \& Viegas,
S. M. M. 2000, A\&A, 361, 1121

\bibitem[Meyer \etal 1998]{m98} Meyer, D. M., Jura, M., \& Cardelli, 
J. A. 1998, ApJ, 493, 222

\bibitem[Meyer 2001]{m01} Meyer, D. M. 2001, XVIIth IAP Colloquium, 
Gaseous Matter in Galaxies and Intergalactic Space, 19-23 June 2001, 
edited by R. Ferlet \etal

\bibitem[Moos \etal 2002]{moos02} Moos, H. W., \etal 2002, ApJS,
140, 3

\bibitem[O'Meara \etal 2001]{om01} O'Meara, J. M., Tytler, D., Kirkman, 
D., Suzuki, N., Prochaska, J. X., Lubin, D., \& Wolfe, A. M. 2001, ApJ, 
552, 718 

\bibitem[Pettini \& Bowen 2001]{pb01} Pettini, M. \& Bowen, D. V. 2001, 
ApJ, 560, 41 

\bibitem[Sonneborn \etal 2002]{s02} Sonneborn, G., \etal 2002, ApJS, 
140, 51

\bibitem[Wood \etal 2002]{w02} Wood, B. E., \etal 2002, ApJS, submitted 
140, 91

\end{thebibliography}
\end{document}